\documentclass[twocolumn,secnumarabic,amssymb, nobibnotes, aps, prl, superscriptaddress]{revtex4}

\usepackage{graphicx}
\usepackage{dcolumn}
\usepackage{bm}
\usepackage{amsmath}
\usepackage{threeparttable}
\usepackage{color} 
\usepackage{epstopdf}
\usepackage[normalem]{ulem}

\begin{document}


\title{Electron-hole tunneling revealed by quantum oscillations in the nodal-line semimetal HfSiS}

\author{M.~R.~van~Delft}
\affiliation{High Field Magnet Laboratory (HFML-EMFL), Radboud University, Toernooiveld 7, Nijmegen 6525 ED, Netherlands.}
\affiliation{Radboud University, Institute for Molecules and Materials, Nijmegen 6525 AJ, Netherlands.}

\author{S.~Pezzini}
\affiliation{High Field Magnet Laboratory (HFML-EMFL), Radboud University, Toernooiveld 7, Nijmegen 6525 ED, Netherlands.}
\affiliation{Radboud University, Institute for Molecules and Materials, Nijmegen 6525 AJ, Netherlands.}

\author{T.~Khouri}
\affiliation{High Field Magnet Laboratory (HFML-EMFL), Radboud University, Toernooiveld 7, Nijmegen 6525 ED, Netherlands.}
\affiliation{Radboud University, Institute for Molecules and Materials, Nijmegen 6525 AJ, Netherlands.}

\author{C.~S.~A. M\"{u}ller}
\affiliation{High Field Magnet Laboratory (HFML-EMFL), Radboud University, Toernooiveld 7, Nijmegen 6525 ED, Netherlands.}
\affiliation{Radboud University, Institute for Molecules and Materials, Nijmegen 6525 AJ, Netherlands.}

\author{M.~Breitkreiz}
\affiliation{Instituut-Lorentz, Universiteit Leiden, P.O. Box 9506, 2300 RA Leiden, The Netherlands}
\affiliation{Dahlem Center for Complex Quantum Systems and Fachbereich Physik, Freie Universit\" at Berlin, 14195 Berlin, Germany}

\author{L.~M.~Schoop}
\affiliation{Department of Chemistry, Princeton University, Princeton, New Jersey 08544, USA}


\author{A.~Carrington}
\affiliation{H. H. Wills Physics Laboratory, University of Bristol, Tyndall Avenue, Bristol BS8 1TL, UK.}

\author{N.~E.~Hussey}
\affiliation{High Field Magnet Laboratory (HFML-EMFL), Radboud University, Toernooiveld 7, Nijmegen 6525 ED, Netherlands.}
\affiliation{Radboud University, Institute for Molecules and Materials, Nijmegen 6525 AJ, Netherlands.}

\author{S.~Wiedmann}
\email{steffen.wiedmann@ru.nl}
\affiliation{High Field Magnet Laboratory (HFML-EMFL), Radboud University, Toernooiveld 7, Nijmegen 6525 ED, Netherlands.}
\affiliation{Radboud University, Institute for Molecules and Materials, Nijmegen 6525 AJ, Netherlands.}

\date{\today}

\begin{abstract}
We report a study of quantum oscillations in the high-field magneto-resistance of the nodal-line semimetal HfSiS. In the presence of a magnetic field up to 31~T parallel to the $c$-axis, we observe quantum oscillations originating both from orbits of individual electron and hole pockets, and from magnetic breakdown between these pockets. In particular, we find an oscillation associated with a breakdown orbit enclosing one electron and one hole pocket in the form of a `figure of eight'. This observation represents an experimental confirmation of the momentum space analog of Klein tunneling. When the $c$-axis and the magnetic field are misaligned with respect to one another, this oscillation rapidly decreases in intensity. Finally, we extract the cyclotron masses from the temperature dependence of the oscillations, and find that the mass of the 'figure of eight' orbit corresponds to the sum of the individual pockets, consistent with theoretical predictions for Klein tunneling in topological semimetals.
\end{abstract}


\maketitle

The advent of topological semimetals (TSMs) created new opportunities to search for relativistic quasi-particles and novel macroscopic quantum phenomena \cite{Burkov2016,Armitage2018}. Two examples of TSMs are Dirac semimetals, which possess fourfold degenerate band crossings in momentum space, and Weyl semimetals (WSM), in which the spin degeneracy is lifted. In the latter case, the band crossings are referred to as Weyl points. These points have opposite chirality leading to various phenomena such as the so-called chiral anomaly in the bulk and Fermi arcs on the surface \cite{Burkov2016,Armitage2018,Wan2011,Potter2014,Nielsen1983,Moll2016}. Within the group of WSMs, there is a class of materials referred to as type-II WSMs with a band structure in which pairs of electron and hole pockets coexist over a range of energy with a topologically protected crossing. For these materials, a new type of quantum oscillation has been theoretically predicted, that results from tunneling between adjacent electron and hole pockets which has been termed 'momentum space Klein tunneling' \cite{O’Brien2016}. Until now, this particular effect has been observed neither in type-II WSMs nor in other classes of TSMs. 

In this Letter, we report on quantum oscillations in the nodal-line semimetal (NLSM) HfSiS that demonstrate the first experimental confirmation of the momentum-space analog of Klein tunneling, the phenomenon in which an electron has a perfect transmission in a $p$-$n$ junction in real space \cite{Katsnelson2006}. It is identified by the characteristic frequency $f_{\beta} - f_{\alpha}$ of the `figure of eight' breakthrough orbit that encloses one electron pocket $\beta$ and one hole pocket $\alpha$. This orbit appears with increasing magnetic field, is only present if $\textbf{B} \parallel c$ and has a cyclotron mass which is the sum of the individual masses of the  two pockets in agreement with theoretical predictions. By further increasing the magnetic field, we find high frequency oscillations that correspond to tunneling across multiple energy gaps as observed in ZrSiS \cite{Pezzini2018}. These oscillations form a complex frequency spectrum that will be briefly discussed towards the end of this Letter.

Nodal-line semimetals are materials in which conduction and valence bands touch each other along a closed trajectory inside the Brillouin zone instead of at discrete nodes in momentum space \cite{Fang2016}. They are of particular interest since screening of the Coulomb interaction is predicted to remain long-ranged and to be weaker compared to conventional metals due to their low density of states at the Fermi level $E_F$ \cite{Huh2016}. In combination with their metallic nature, this renders NLSMs more susceptible to different types of order such as superconductivity, magnetism or charge order \cite{Liu2016,Roy2017}. Indeed, recent experiments on ZrSiS have shown evidence for electronic correlations \cite{Pezzini2018} and a superconducting phase induced by a non-superconducting metallic tip \cite{Aggarwal2018}. 

Experimentally verified material systems belonging to the class of NLSMs include PbTaSe$_2$ \cite{Bian2016}, PtSn$_4$ \cite{Wu2016} and compounds with a general formula of XSiY (X = Zr, Hf and Y = S, Se, Te) that are of particular interest as they reveal a number of distinct features predominantly investigated in the Dirac NLSM ZrSiS \cite{Neupane2016,Schoop2016,Wang2016,Ali2016,Hu2016,Singha2016,Hu2016_2,Topp2016}. Firstly, in the absence of spin-orbit coupling (SOC), \textit{all} bands nearby the Fermi energy $E_F$ in ZrSiS have a Dirac-like dispersion forming a line node in the Brillouin zone giving rise to a cage-like Fermi surface \cite{Schoop2016}. Around 0.7~eV below $E_F$ at the X-point in the Brillouin zone, there is a linear band crossing associated with 2D Dirac fermions which is protected by non-symmorphic symmetry and survives in the presence of strong spin-orbit coupling (SOC) \cite{Young2015}. Secondly, surface states are observed in angle-resolved photo-emission spectroscopy experiments in ZrSiS and attributed to a reduced symmetry at the surface in contrast to other topological materials with band inversion \cite{Topp2017}. 

Until now, HfSiS has been much less investigated than other members of the ZrSiS family \cite{Takane2016,Chen2017,Kumar2017}. As with all the compounds with a general formula of XSiY, HfSiS has a PbFCl-type structure (tetragonal space group P4/nmm) and layers of Hf and S are sandwiched between Si square nets extending in the $ab$-plane. Its electronic band structure, calculated by Density Functional Theory (DFT), is displayed in Fig.~S1 in the Supplemental Material \cite{SI} and shows linearly dispersive bands over a large energy range ($\sim$~2~eV around $E_F$). A three-dimensional representation of the Fermi surface is presented in Fig.~S2 in the Supplemental Material \cite{SI}. Due to the enhanced spin-orbit coupling in HfSiS, the calculated size of the band gap in the Dirac spectrum is 2 - 3 times larger than found in ZrSiS \cite{Schilling2017}. Indeed, the size of this gap is such that tunneling between adjacent orbits is not expected up to very high magnetic fields.

\begin{figure}
  \centering
	\includegraphics[width=0.4\textwidth]{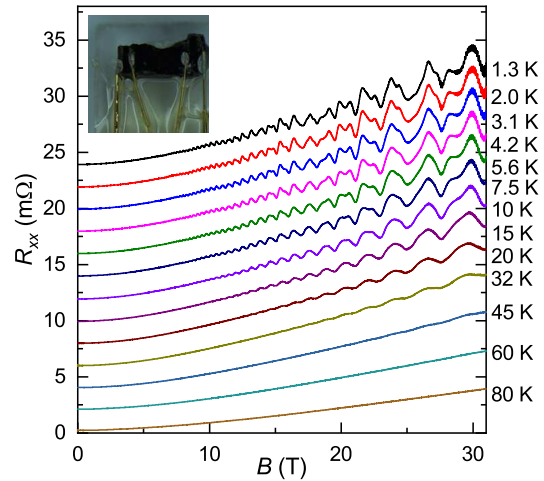}
	\vspace{-0.5cm}
	\caption{Longitudinal resistance $R_{xx}$ as a function of the magnetic field for temperatures between 1.3 and 80~K up to 31~T with $\textbf{B}\parallel c$. Each successive curve is offset by 2~m$\Omega$ for clarity. The inset shows an image of a typical HfSiS single crystal with attached current and voltage probes.}
	\label{fig1}
\end{figure}

Figure~\ref{fig1} shows a series of resistance sweeps as a function of the magnetic field up to 31 T on a HfSiS single crystal at different temperatures between 1.3 and 80~K. The magnetic field is applied parallel to the $c$-axis. As seen in the figure, HfSiS displays Shubnikov-de Haas (SdH) oscillations with multiple frequencies superimposed on a positive magneto-resistance (MR). Above 15~T, additional high frequency oscillations are observed which we shall return to later.

\begin{figure}
  \centering
	\includegraphics[width=0.5\textwidth]{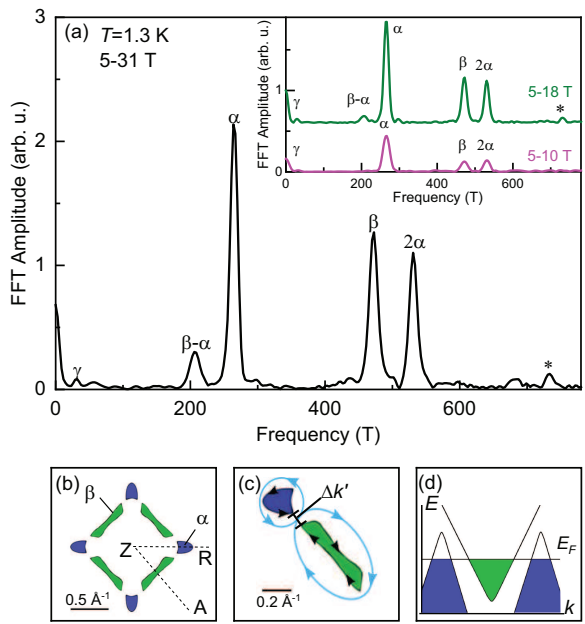}
	\vspace{-3.2cm}
	\caption{Fast Fourier transform analysis of the data from 5 to 31~T at $T$ = 1.3~K. (a) Low-frequency FFT spectrum for different field ranges (inset). The FFT peaks with the individual extremal orbits are labeled as $\alpha$, $\beta$ and $\gamma$. The `figure of eight'-orbit with the frequency $\beta - \alpha$ is also indicated. (b) Sketch of the projection of the diamond-shaped Fermi surface in the Z-R-A plane in the first Brillouin zone for $\textbf{B}\parallel c$ ($\Theta =$ 0$^{\circ}$). The gap $\Delta k'$ between the orbits is exaggerated here for illustration. The electron pockets, $\beta$ are indicated in green; the hole pockets, $\alpha$ in blue. (c) Expanded view that highlights the `figure of eight'-orbit in momentum space created by an individual electron and hole orbit. The arrows indicate the direction with which charge carriers can move around the pockets and $\Delta k'$ represents the separation in $k$-space between the adjacent pockets. (d) Schematic of the dispersion relation $E$($k$). The horizontal line indicates the Fermi level $E_F$.}
	\label{fig2}
\end{figure}

In Fig.~\ref{fig2}, we perform a Fast Fourier Transform (FFT) analysis of the 1.3 K sweep in Fig.~\ref{fig1}. Fig.~\ref{fig2}(a) illustrates the FFT amplitude in the low-frequency part of our spectrum that has been generated from data between 5 and 31~T. The observed frequencies, $f$, in the FFT spectra are related to the extremal area $A_F$ of the individual pockets via the Onsager relation $f = (\hbar/2\pi e)\cdot A_F$ \cite{Shoenberg1984}. 

In this figure, a number of clearly resolved peaks are visible. The two largest amplitude peaks at 264~T and 473~T we label as $\alpha$ and $\beta$ respectively, and several other peaks can be identified as the second harmonic of $\alpha$ or the sum and difference of $\alpha$ and $\beta$. $f_{\alpha}$ was reported previously \cite{Kumar2017}. The origin of these orbits can be deduced by comparing the experimental variation of the frequencies with magnetic field angle ($\theta$) as $B$ is rotated from $c$ towards $a$, to the DFT calculations (see Supplemental Material Fig.~S3 \cite{SI}). The $\alpha$ orbit is close to the  predictions for the hole pocket located at the vertex of the diamond-shaped Fermi surface in the Z-R-A plane and is highlighted in blue in Fig.~\ref{fig2}(b). The $\beta$ orbit has a very similar $\theta$ dependence (including the splitting for $\theta>20^\circ$) to that predicted for the electron pocket running parallel to the top rung of the nodal loop and indicated in green in Fig.~\ref{fig2}(b). The absolute frequency $f_{\beta}$ is experimentally $\sim 25\%$ lower than predicted.

Additionally, we observe a low-frequency orbit at 32~T labeled $\gamma$ (also observed in Ref.~\cite{Kumar2017}), which we attribute to a closed orbit in a plane between the Z-R-A and the $\Gamma$-M-X planes. The peak at 734~T, marked with an asterisk in Fig.~\ref{fig2}a, matches the sum of $\beta + \alpha$, which semi-classically should be `forbidden' because it requires the sudden reversal of the electron momentum \cite{footnote1}.

The main finding of this work is the observation of a peak in the FFT spectrum at a frequency of 207~T, which is consistent with the assignment $\beta-\alpha$. An orbit that includes these two pockets is semi-classically allowed and leads to quantum oscillations that are periodic in 1/$B$ with a frequency set by the difference between the two contributing pockets \cite{Kaganov1983,O’Brien2016,Alexandradinata2017,Alexandradinata_1_2017}. In the context of TSMs, such an orbit has been predicted to appear in type-II Weyl semimetals \cite{O’Brien2016}. This orbit is only possible due to momentum-space Klein tunneling between two closely spaced pockets of opposite character in which the different circulation paths of the carriers lead to a reduced Aharonov-Bohm phase. Fig.~\ref{fig2}(c) shows a sketch of the orbit, which we refer to as the `figure of eight' orbit, in the Z-R-A-plane in momentum space. The corresponding DFT calculation in this plane and the breakdown gap are illustrated in Fig.~S4 in the Supplemental Material \cite{SI}. As shown in the inset of Fig.~\ref{fig2}(a), where we present FFT spectra calculated for different magnetic field ranges, the $\beta - \alpha$-orbit is absent at low fields. This is consistent with the picture of magnetic breakdown \cite{Cohen1961, Kaganov1983}, where tunneling of charge carriers is only possible between adjacent orbits in the Z-R-A-plane once the magnetic field is larger than a certain breakdown field.

\begin{table*}[ht]
	\caption{Observed frequencies and extracted cyclotron masses in the field range from 10 to 31 T from the FFT analysis \cite{FFT} in comparison to DFT calculations.} 
	\centering 
	\begin{tabular}{c | c | c | c | c | c} 
		\hline 
		Orbit & FFT analysis & DFT prediction & Plane & $m_c$/$m_e$ (exp.)\tnote{a} & $m_c$/$m_e$ (DFT)\\ 			
		\hline
		
		\hline 
		$\alpha$ 					& 264 T & 294 T  & Z-R-A		&  (0.177 $\pm$ 0.003) & 0.17 \\ 
		$\beta$  					& 473 T & 631 T  & Z-R-A 		& (0.48 $\pm$ 0.02)    & 0.63  \\ 
		$\gamma$ 					& 32 T  & 38 T  &  in between 	& (0.17 $\pm$ 0.06) \\ 
		$2\alpha$ 					& 528 T &   -    & Z-R-A 		& (0.32 $\pm$ 0.02)  \\
		$\beta - \alpha$ 			& 207 T &   -    & Z-R-A 		& (0.59 $\pm$ 0.06) \\
		$\beta + \alpha$  			& 734 T &   -    &          	& (0.22 $\pm$ 0.08)  \\
		\hline
	\end{tabular}
	\label{table_frequencies}
\end{table*}

\begin{figure}
  \centering
	\includegraphics[width=0.4\textwidth]{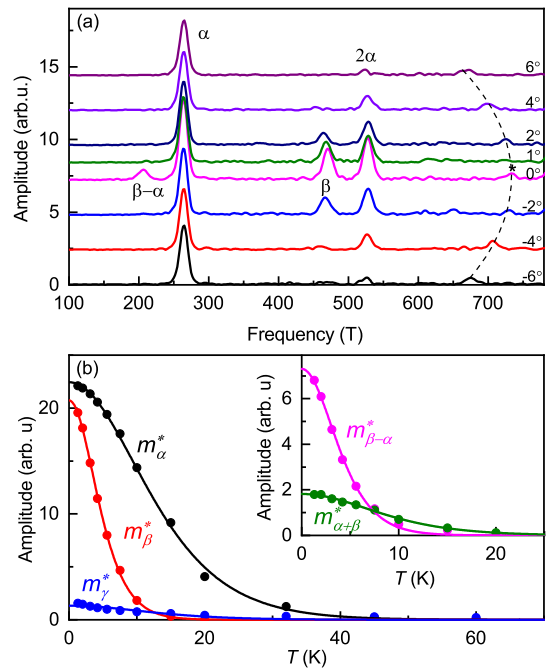}
	\caption{(a) FFT spectra at $\textbf{B}\parallel c$ ($\Theta =$ 0$^{\circ}$) and at small tilt angles for $T$=1.3~K. While the individual orbits disperse differently with increasing tilt angle, the $\beta - \alpha$ orbit can only be observed at $\Theta =$ 0$^{\circ}$. The dashed line is a guide to the eye, following the $\alpha+\beta$, ($*$), frequency. (b) Mass plot for the individual orbits $\alpha$ and $\beta$, and the $\beta - \alpha$ and $\alpha+\beta$ orbits (inset). The resulting cyclotron masses are specified in Table~\ref{table_frequencies}.}
	\label{fig3}
\end{figure}

We focus now on Fig.~\ref{fig3} which illustrates the angle and temperature dependence of the individual $\alpha$ and $\beta$ orbits as well as the $\beta - \alpha$ and $\beta + \alpha$ ($*$) orbits. Fig.~\ref{fig3}(a) shows the angle dependence at 1.3~K for small angles close to $\textbf{B}\parallel c$. Interestingly, the $\beta - \alpha$ orbit vanishes within two degrees of misalignment between the magnetic field and the $c$-axis. This is unexpected, as the breakdown gap $k_g$ would be expected to grow only very weakly with misalignment angle. Moreover, the different pockets of the Fermi surface $\alpha$ and $\beta$ disperse in different ways once the orientation of $B$ is changed with respect to the $c$-axis. $f_{\alpha}$ is almost constant in this range of angles, whereas $f_{\beta}$ changes slightly with tilt angle and disappears between 4$^{\circ}$ and 8$^{\circ}$ corresponding to a so-called spin zero where different spin split Fermi surfaces interfere. In contrast to the $\beta- \alpha$ orbit, the orbit with the `forbidden' frequency $\alpha + \beta$ disperses strongly with angle which could be associated with a breakdown orbit \cite{Eddy1982, Meyer1995} or magnetic interaction \cite{Shoenberg1984,SI}. A full angle dependence where the evolution of the main orbits is shown together with expectations from the DFT calculations can be found in Fig.~S3 in the Supplemental Material \cite{SI}.

Fig.~\ref{fig3}(b) shows the evolution of the FFT amplitude for the individual orbits $\alpha$, $\beta$ and $\gamma$ with temperature. Using the standard Lifshitz-Kosevich formula for the thermal damping of the oscillations \cite{Shoenberg1984,Lifshitz1953}, we extract the cyclotron masses $m^* = \frac{\hbar^2}{2\pi}\frac{d}{dE}A$ for each of the observed orbits, where $A$ is the area enclosed by the orbit in momentum space. The results are shown in Table \ref{table_frequencies}, with $m_e$ the free electron mass. The error margins in these results are estimated from the noise level and interference from nearby peaks or side-lobes of the FFT window function while the error from the fit function is an order of magnitude smaller. Although the cyclotron mass for the $\alpha$-pocket is in agreement with the DFT prediction, the cyclotron mass of the $\beta$-pocket is found to be smaller than predicted, however, the discrepancy is almost perfectly proportional to the DFT overestimate of the frequency.

For the `figure of eight' orbit, one obtains $|m^*_{\beta-\alpha}| = |m^*_\beta|+|m^*_\alpha|$ \cite{SI}. Within our experimental uncertainty, we find that $m^*_{\beta-\alpha}$ indeed agrees with the sum of the cyclotron masses from the individual pockets involved in this breakdown orbit. These findings, the frequency $\beta - \alpha$ in the FFT and the extracted mass $m^*_{\beta-\alpha}$ are consistent with theoretical predictions for Klein tunneling \cite{O’Brien2016,Kaganov1983}. Assuming that the classical forbidden orbit $\beta + \alpha$ is indeed formed by two pockets with electron and hole like character, $m^*_{\beta+\alpha} = |m^*_\beta|-|m^*_\alpha|$ which in good agreement with our observation. However, this would require a mechanism reversing the direction of circulation of carriers around a closed orbit while preserving the sign of $dA\big/dE$. Despite the agreement concerning $m^*_{\beta+\alpha}$, we emphasize that the dispersion of $\beta + \alpha$ ($*$) shown in Fig.~\ref{fig3}(a) is different from the angle dependence of the individual $\alpha$ and $\beta$ pockets. Thus, the origin of the $\beta + \alpha$-orbit demands further investigation \cite{footnote2}.

\begin{figure}
  \centering
	\includegraphics[width=0.48\textwidth]{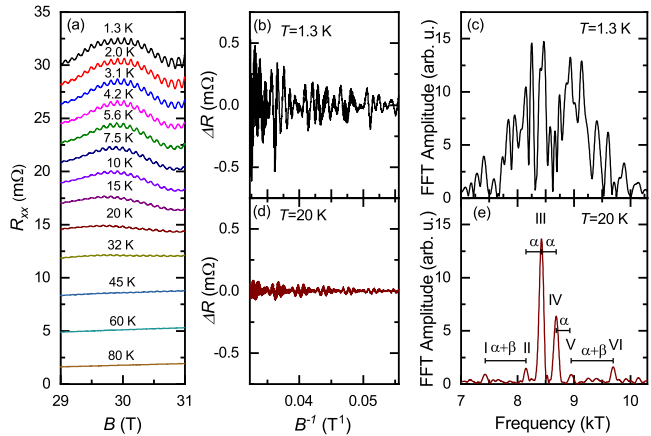}
	\caption{High-frequency quantum oscillations. (a) $R_{xx}$ as a function of $B$ between 29 and 31~T for fixed temperatures ranging from 1.3 to 80~K ($\textbf{B}\parallel c$). The curves are offset for clarity by 2~m$\Omega$. (b,c) and (d,e) Oscillatory resistance $\Delta R$ versus 1/$B$ and FFT analysis for 1.3 and 20~K, respectively. The spacer markers for the 20 K FFT indicate that the frequencies are separated by different combinations of the $\alpha$ and $\beta$ frequencies.}
	\label{fig4}
\end{figure}

Finally, we turn to the high frequency oscillations that appear above 15~T in Fig.~\ref{fig1}. Figure~\ref{fig4}(a) shows an enlarged view of these oscillations between 29 and 31~T for different temperatures. By removing the background that originates from the MR as well as from the low-frequency oscillations discussed so far, we find a complex oscillatory pattern in $\Delta R$ as a function of the inverse magnetic field, $B^{-1}$, shown for 1.3 and 20~K in Figs.~\ref{fig4}(b) and (d). The corresponding spectra of the FFT then comprise a series of peaks ranging from 7 to around 10~kT with the most pronounced ones around 8 and 9~kT. There are no individual orbits in the DFT derived Fermi surface with areas that match such high frequencies. Thus, we attribute these peaks in the FFT spectrum to magnetic breakdown orbits that encircle the diamond in the Z-R-A-plane, see Fig.~\ref{fig2}(b), which are also strongly suppressed for small tilt angles away from the $c$-axis as found in ZrSiS \cite{Pezzini2018},

Unfortunately, the resolution of the low-temperature FFT spectra of this high frequency part of our data does not allow us to investigate individual peaks in detail as their close spacing leads to a considerable overlap of adjacent peaks. At 20~K, however, most of the frequencies have vanished and only six peaks remain in the spectrum, see Fig.~\ref{fig4}(e). Two of these peaks (III and IV) have an amplitude that is almost as large as at 1.3~K, while the other four peaks are significantly damped. This also precludes our ability to investigate the $T$-dependence of the quantum oscillation amplitude that in ZrSiS showed anomaluous behaviour \cite{Pezzini2018} and requires further detailed investigation. The peaks II-V are all spaced in frequency by 264 T (i.e. the frequency of the orbit around the $\alpha$-pocket), whereas the peaks I and VI are a distance of approximately 734~T ($\alpha+\beta$) away. This suggests that these peaks correspond to orbits around the Fermi surface that incorporate different multiples of the $\alpha$ and $\beta$ pockets.

In conclusion, we have demonstrated the first experimental confirmation of Klein tunneling in momentum space in the NLSM HfSiS. This special type of magnetic breakdown is only present if the magnetic field is oriented precisely parallel to the $c$-axis and its cyclotron mass is found to be the sum of the masses of the individual electron and hole pockets. Although breakdown has been extensively studied in simple elements, and in organic metals, this special type of magnetic breakdown between electron and hole pockets has not been observed previously. The reason is that in compensated organic metals the Fermi surface is always strongly two dimensional, so that the areas of the electron and hole pockets are exactly equal at all $k_z$, hence the difference frequency is always zero and unobservable. The observation in the NLSM HfSiS implies that the phenomenon of momentum space Klein tunneling is not confined to type-II WSMs. Our results suggest, however, that this is a generic property of semimetals with adjacent electron and hole pockets.

\begin{acknowledgments}
This work was supported by the High Field Magnet Laboratory - Radboud University/Foundation for Fundamental Research on Matter (HMFL-RU/FOM) — a member of the European Magnetic Field Laboratory and by the UK Engineering and Physical Sciences Research Council (Grant No. EP/R011141/1). We thank Tom O'Brien, Stephen Hayden and Mark Kartsovnik for helpful discussions.
\end{acknowledgments}


\begin{thebibliography}{39}
\bibitem{FFT}
The experimental masses were determined by fitting the amplitude of the FFT peak versus temperature assuming an effective average field equal to the average inverse field in the window. A more refined procedure, taking into account the finite Dingle term, would increase the $\alpha$ and $\beta$ masses by 8\% and 15\% respectively.
\end{thebibliography}
\end{document}